\documentclass[doublespacing]{elsart}

\usepackage{graphicx}
\usepackage{amssymb}
\begin{document}

\begin{frontmatter}

\title{Multi-particle correlation function to study short-lived nuclei}




\author[laval]{F.~Grenier}
\author[ganil]{A.~Chbihi}
\author[laval]{R.~Roy}
\author[infn]{G.~Verde}
\author[laval]{D. Th\'{e}riault}
\author[ganil]{J.D.~Frankland}
\author[ganil]{J.P.~Wieleczko}
\author[orsay]{B.~Borderie}
\author[lpc]{R.~Bougault}
\author[saclay]{R.~Dayras}
\author[orsay,cnam]{E.~Galichet}
\author[ipn,orsay]{D.~Guinet}
\author[ipn]{P.~Lautesse}
\author[orsay]{N.~Le~Neindre}
\author[lpc]{O.~Lopez}
\author[ganil]{J.~Moisan}
\author[saclay]{L.~Nalpas}
\author[lpc]{P.~Napolitani}
\author[nipne,lpc]{M.~P\^arlog}
\author[orsay]{M.~F.~Rivet} 
\author[napoli]{E.~Rosato}
\author[lpc]{B.~Tamain}
\author[lpc]{E.~Vient}
\author[napoli]{M.~Vigilante}
\author{(INDRA collaboration)}
\address[laval]{ Laboratoire de Physique Nucl\'eaire, Universit\'e Laval,
Qu\'ebec, Canada G1K 7P4.}
\address[ganil]{ GANIL, CEA et IN2P3-CNRS, B.P.~5027, F-14076 Caen Cedex, France.}
\address[infn]{Istituto Nazionale di Fisica Nucleare, Sezione di Catania, 
64 Via Santa Sofia, I-95123, Catania, Italy}
\address[orsay]{ Institut de Physique Nucl\'eaire, IN2P3-CNRS, F-91406 
Orsay Cedex,  France.}
\address[lpc]{ LPC Caen, ENSICAEN, Universit\'e de Caen, CNRS-IN2P3, Caen, France.}
\address[saclay]{ DAPNIA/SPhN, CEA/Saclay, F-91191 Gif sur Yvette, France.}
\address[cnam]{ Conservatoire National des Arts et M\'etiers, F-75141 Paris
Cedex 03.}
\address[ipn]{ Institut de Physique Nucl\'eaire, IN2P3-CNRS et Universit\'e de Lyon, Universit\'e Claude Bernard,
F-69622 Villeurbanne, France.}
\address[nipne]{ National Institute for Physics and Nuclear Engineering, RO-76900
Bucharest-M\u{a}gurele, Romania.}
\address[napoli]{ Dipartimento di Scienze Fisiche e Sezione INFN, Univ. di 
Napoli ``Federico II'', I-80126 Napoli, Italy.}

%%%%%%%%%%%%%%%%%%%%%%%%%
%%%%%%%%%%%%%%%%%%%%%%%%%

\begin{abstract}
Unstable $^{10}$C nuclei are produced as quasi-projectiles in $^{12}$C+$^{24}$Mg collisions at E/A = 53 and 95 MeV. The decay of their short-lived states is studied with the INDRA multidetector array  via multi-particle correlation functions. 
The obtained results show that heavy-ion collisions can be used as a tool to access 
spectroscopic information of unbound states in exotic nuclei, such as their energies and the relative importance of different sequential decay widths. 
\end{abstract}

\end{frontmatter}

Energetic heavy-ion collisions have been extensively studied to extract
information about the properties of nuclear matter under extreme conditions \cite{wci06}. 
These studies have also shown that a large variety of isotopes is produced
during the dynamical evolution of the reaction. Some of these isotopes, being 
far from the valley of stability, live temporarily and decay by
particle emission. Their unbound states can then be 
isolated and studied by means of correlation techniques \cite{han52,han56,koo77,cha95,tan04}. 
As an example, proton-$^{7}$Be correlation functions have been recently measured 
in central Xe+Au collisions at E/A=50 MeV \cite{tan04} to determine the spin of internal unbound states of 
the astrophysically important $^{8}$B nucleus. In this respect, a collision between two heavy 
ions can be viewed not only as a tool to study nuclear dynamics but also as a laboratory to produce several
nuclear species in one single experiment and study their spectroscopic
properties. This aspect of heavy-ion collision experiments represents an important perspective 
to access information about the properties of very exotic nuclei. 
Among all exotic nuclear species that can be produced in nuclear reactions, $^{10}$C 
can be considered as an especially interesting one~\cite{char07,cur08}.  An extended Fermionic molecular dynamics (FMD) approach \cite{fed90} allows one to describe light nuclei with cluster and halo structures \cite{nif04}. Within the context of similar approaches, Antisymmetrized molecular dynamics (AMD) calculations predict a molecular structures for the ground state of $^{10}$C \cite{kan97}. Exotic features, such as molecular states and 
clustering \cite{cha88a,cha88b,wuo92,dip99,yam05,fre99,fre01,fre06,voertzen06}, have indeed attracted the interest of a 
large community and particle correlation analyses have been used to access such features 
experimentally \cite{wuo92,dip99,yam05,fre99,fre01,fre06,voertzen06}. However, previous studies of $^{10}$C nuclei have 
hardly  reconstructed those states lying above the particle emission
threshold of 3.73 MeV \cite{niveau8}. These states decay into a single final configuration 
constituted by two alpha particles and two protons (2$\alpha$+2$p$). All intermediate states that can be 
formed starting from the decay of $^{10}$C nuclei by charged particle emission (i.e. $^{2}$He, $^{5}$Li, $^{6}$Be, $^{8}$Be, $^{9}$B) are unstable. Reconstructing the decay paths of these $^{10}$C states therefore requires the detection of all intermediate and final particles with high energy and angular resolution. Furthermore, a large solid angle detector coverage is required to completely detect all decay products, due to the excitation energy of these high lying states. 

In this article we study  $^{10}$C nuclei produced as excited quasi-projectiles (QP) in $^{12}$C+$^{24}$Mg 
peripheral collisions at E/A=53 and 95 MeV. The experimental data are collected with the INDRA multidetector array  \cite{pou95,pou96}.   By means of three- and four-particle correlation
functions we provide experimental evidence of sequential decay modes
for excited $^{10}$C states through the production of intermediate unstable $^{9}$B,
$^{6}$Be and $^{8}$Be nuclei. Exploring the relative contributions of these sequential 
decay processes to the total decay widths of $^{10}$C states provides important spectroscopic 
information which is relevant to access branching ratios and spins. 

In the first section, the experimental setup and the event selection criteria are presented. The 
multi-particle correlation analysis techniques used to explore sequential decay modes are 
illustrated in section~\ref{sec-cor} by selecting the special case of $^{12}$C states. The last section of this work is finally devoted to the study of  $^{10}$C and its unbound states decaying into 2$\alpha$+2$p$ .

\section{Experimental setup and event selection}

\begin{figure}
\centering
\includegraphics[scale=0.7]{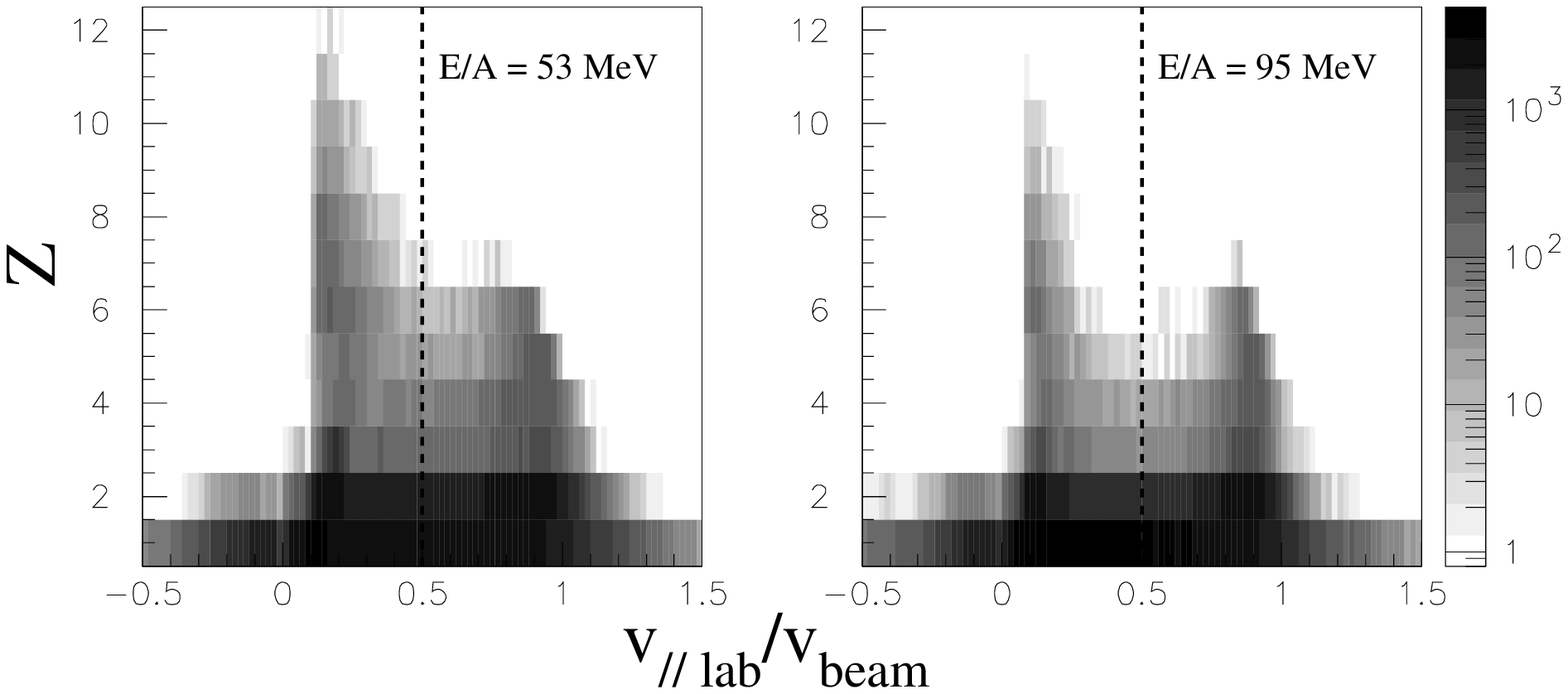}
\caption{Scatter plots of charge (Z) vs velocity in the longitudinal direction, $V_{//lab}$, normalized to the beam velocity, $V_{beam}$. 
The left panel refers to E/A = 53 MeV beam energy; the right panel refers to E/A = 95 MeV beam energy. The dashed line indicates the cut used to select projectile break-up events (see text).}
\label{Zvz}
\end{figure}

\begin{figure}
\centering
\includegraphics[scale=0.7]{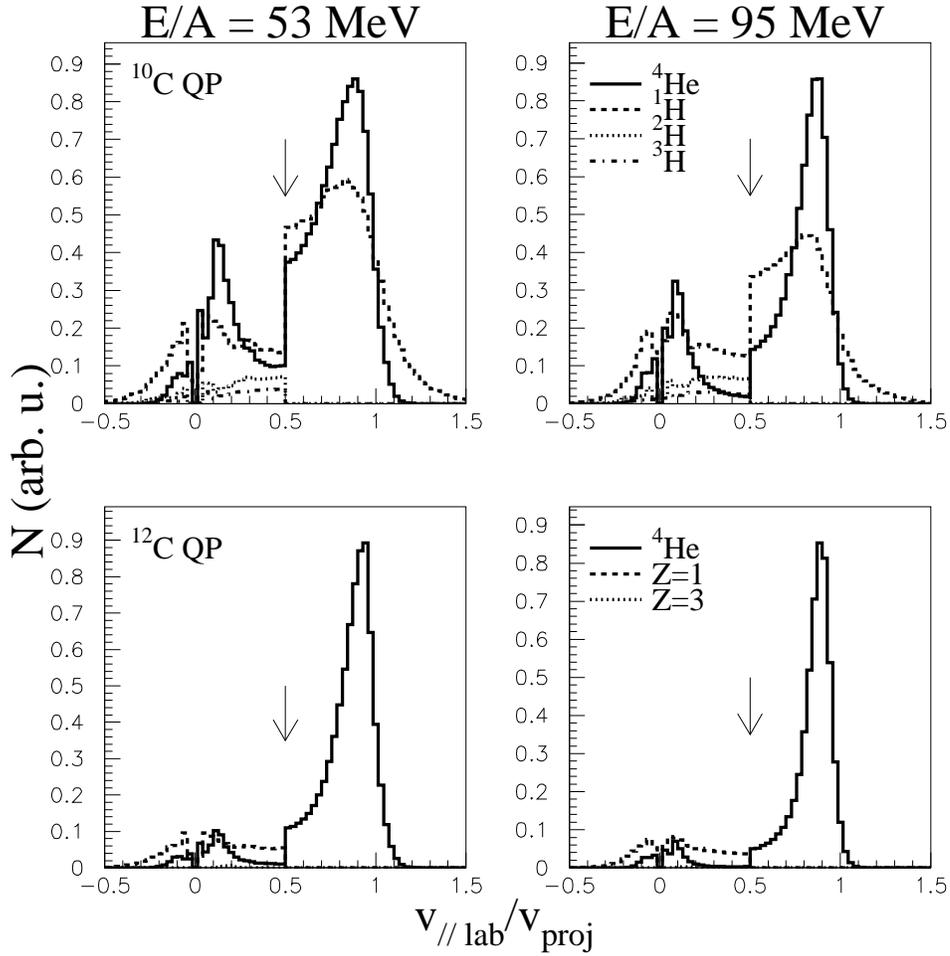}
\caption{Distributions of parallel velocities, $V_{//lab}$, normalized to the beam velocity, $V_{proj}$.
for the $^{12}$C QP decay (bottom panel) and for the $^{10}$C QP decay (top panel). The case of E/A = 53 MeV beam energy is shown on the 
left panels and the case of E/A = 95 MeV beam energy is shown on the right panels. The arrows show where the event selection cut
was applied.}
\label{vpar}
\end{figure}

A 2 mg/cm$^{2}$-thick $^{24}$Mg target was bombarded by beams of $^{12}$C at E/A=53 and 95 MeV, with
an intensity of a few 10$^{7}$ pps, produced by the GANIL cyclotrons. The data presented in this 
work were collected with the INDRA multidetector array \cite{pou95,pou96}. The data acquisition 
trigger required an event multiplicity greater than 2. This trigger condition strongly favors peripheral events. 
Fig.~\ref{Zvz} shows the velocity distribution along the beam direction, V$_{//lab}$/V$_{beam}$ (normalized to the velocity of the beam, V$_{beam}$) vs the charge of the detected particles  at E/A=53 (left panel) and 95 MeV (right panel). These distributions show that there are two sources of emission of charges Z$>$1 having parallel velocities 
centered around values of V$_{//lab}\sim$0.2 V$_{beam}$ and V$_{//lab}\sim$0.9 V$_{beam}$, respectively. In order to select excited Carbon (Z=6) QP breakup events, we take particles with V$_{//lab}\geq$V$_{beam}/$2 and we also require that the sum of their charges  is equal to six units.  Among all the selected channels 
of Carbon QP breakup, we focus on those corresponding to the 3-alpha decay of 
$^{12}$C$^{*}$ (i.e. $^{12}$C$^{*}\rightarrow$3$\alpha$) and to the 2$\alpha$-2p decay of 
$^{10}$C$^{*}$ (i.e. $^{10}$C$^{*}\rightarrow$2$\alpha$+2p). V$_{//lab}$/V$_{beam}$ distributions for Z = 1-2 particles in the selected  $^{12}$C$^{*}$ and $^{10}$C$^{*}$ breakup events are shown on 
Fig.~\ref{vpar}. The discontinuity at V$_{//lab}$=V$_{beam}/$2 is due to the selection criteria. 
More details about isolating the QP decay products can be found on Refs.~\cite{Gre06,chb05}.

\begin{figure}
\centering
\includegraphics[scale=0.7]{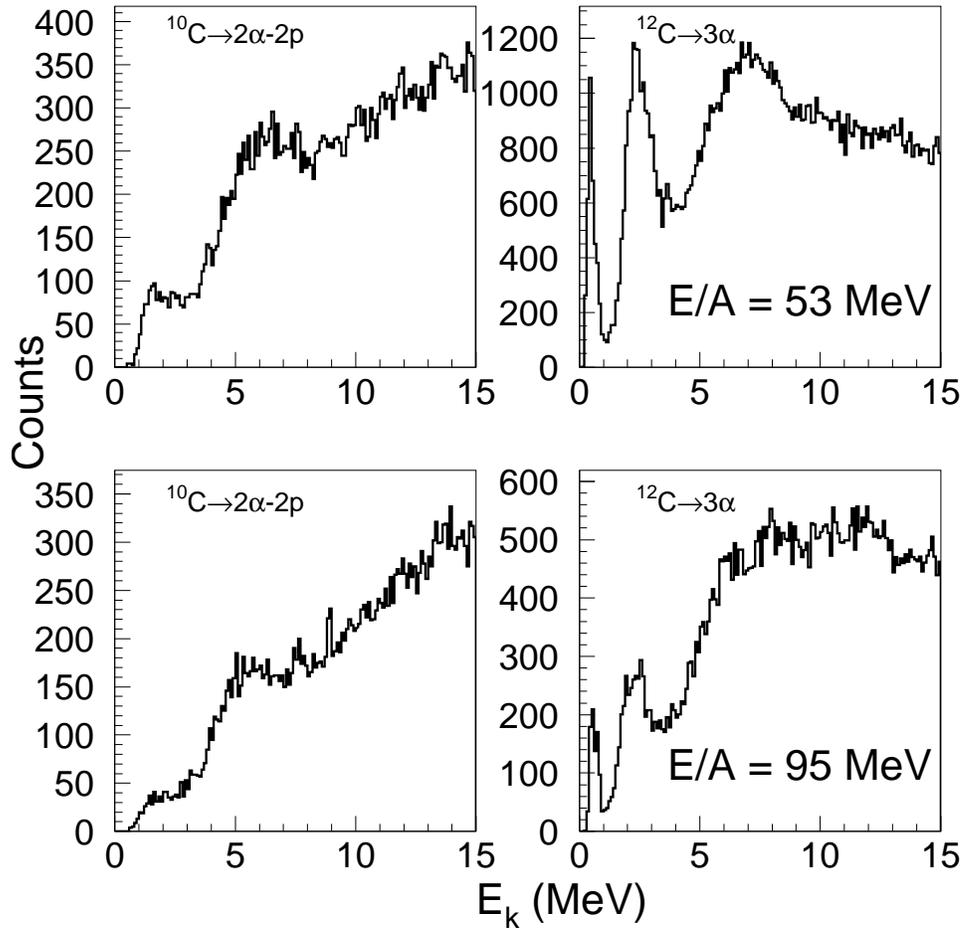}
\caption{E$_{k}$ distributions for the $^{10}$C 
QP channel (left panels) and $^{12}$C QP channel (right panels). Data corresponding to E/A = 53 MeV (95 MeV) beam energies are shown on 
the top (bottom) panels.}
\label{ekdist}
\end{figure}

Fig.~\ref{ekdist} shows distribution of E$_{k}$, defined as the total kinetic energy in the center of mass of the  2$\alpha$-2p (left panels) and 3$\alpha$ (right panels) exit channels.  These distributions display several peaks, more pronounced in the case of  the 3$\alpha$ exit channel at lower incident energies. The position of these peaks does not depend on the incident 
energy. 

\section{Multi-particle correlations}
\label{sec-cor}
\begin{figure}
\centering
\includegraphics[scale=0.7]{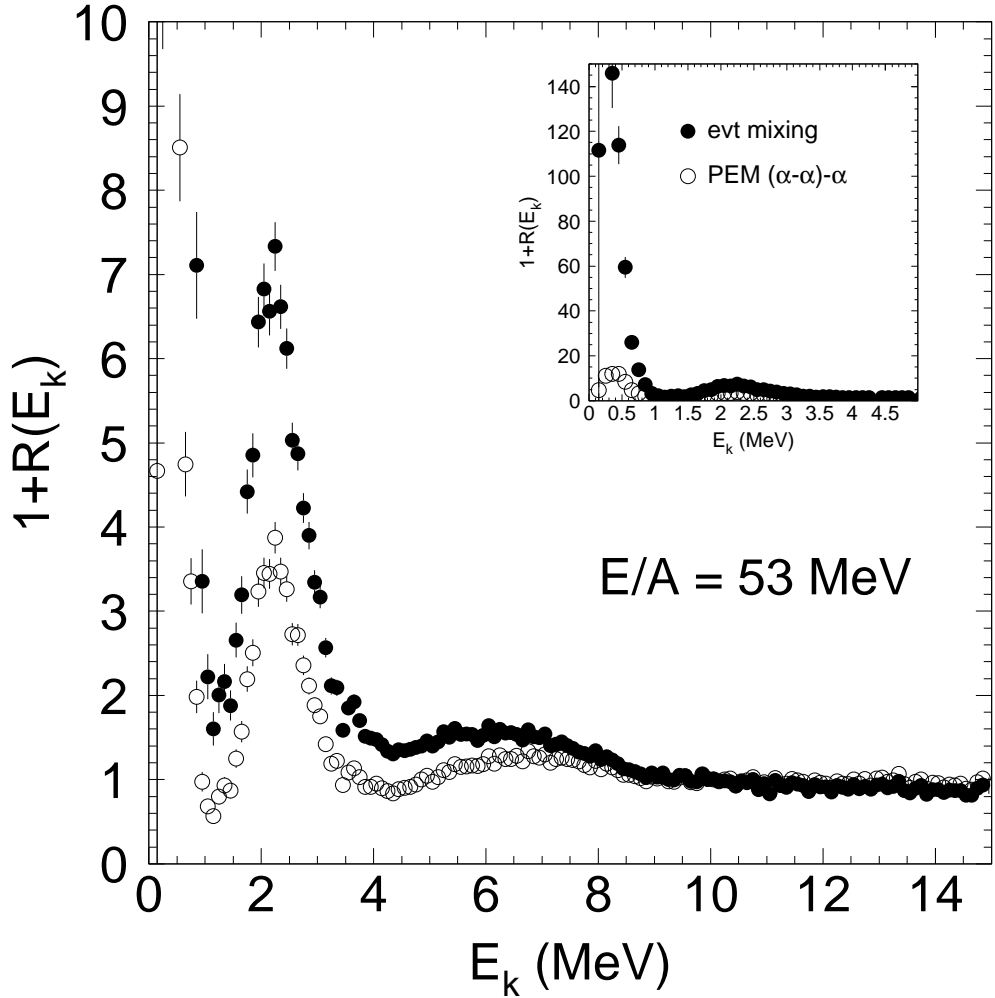}
\caption{Three-$\alpha$ correlation functions. Full dots: 1+R(E$_{k}$) constructed with the standard event mixing technique. 
Open symbols: 1+R(E$_{k}$) constructed with the PEM techniques (see text). An expanded view of the correlation 
function at $E_{k}\leq$5 MeV is shown in the inset. Particles pairs within parentheses are taken from the same event.}
\label{corr12c}
\end{figure}

In order to illustrate our analysis techniques, we first study the most peripheral events where excited $^{12}$C 
QP are produced and are reconstructed by detecting three alpha particles with $V_{//lab}>V_{beam}/2$.
We define the three-alpha particle correlation function, 1+R(E$_{k}$), as
\begin{equation}
1+R\left(E_{k}\right)=\frac{Y_{corr}(E_{k})}{Y_{uncorr}(E_{k})}.
\label{eqfctcorr}
\end{equation}
where the correlated yield spectrum, $Y_{corr}(E_{k})$, is constructed with
3 $\alpha$ particles detected in the same event. This spectrum is sorted with respect to the 
quantity E$_{k}$, already defined in Fig.~\ref{ekdist}. If a $^{12}$C QP is produced at an excitation energy 
$E^{*}$, it follows that $E_{k}=E^{*}+Q$, where 
$Q$ is the mass difference in the $^{12}$C$\rightarrow 3\alpha$ channel. 
The uncorrelated yield spectrum, $Y_{uncorr}(E_{k})$, can be constructed with different techniques. 
In previous works~\cite{cha95} a dedicated Monte-carlo simulation was used to determine 
the uncorrelated background of multi-particle configurations. Here we rather use an event-mixing 
technique ~\cite{Zaj84,dri84}, where the three uncorrelated alpha particles are taken from three different 
events. The full dots on Fig.~\ref{corr12c}
show the resulting 3$\alpha$ correlation function in $^{12}$C+$^{24}$Mg collisions at E/A=53 MeV. 
The Q-value for the 3$\alpha$ decay
of $^{12}$C is -7.27 MeV (see the decay scheme of  $^{12}$C  in Refs.~\cite{niveau11,niveau8}). The first peak observed on Fig. \ref{corr12c} at $E_{k}\simeq$0.5~MeV
corresponds to the 7.65 MeV state in $^{12}$C. The second peak, centered
at $E_{k}\simeq$2.2 MeV, can be associated to the 9.65 MeV state,
while the third bump at $E_{k}\simeq$6 MeV results from
the overlap of highly lying closely packed states. 

\begin{figure}
\centering
\includegraphics[scale=0.7]{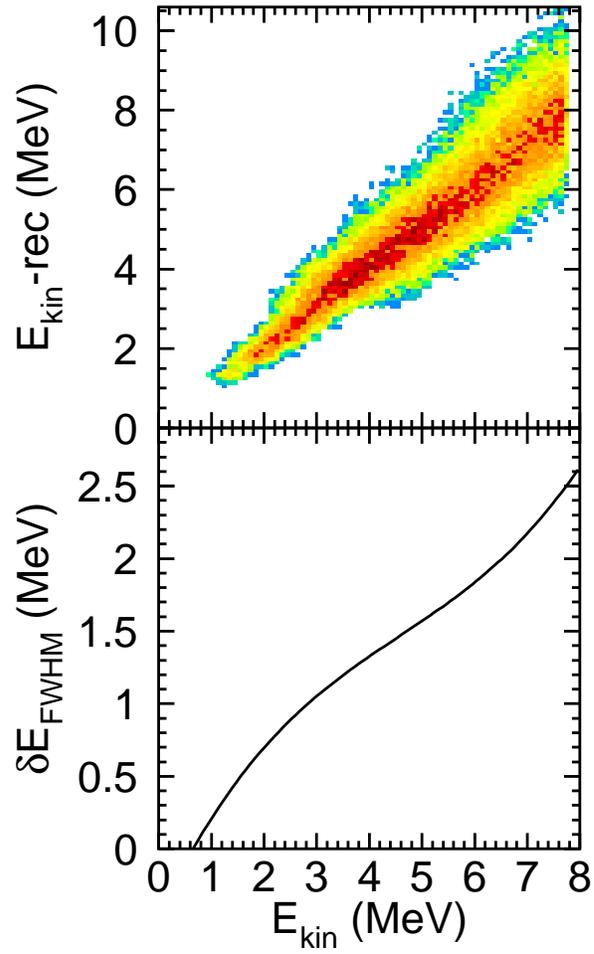}
\caption{Monte Carlo simulation of the breakup of $^{12}$C into three alpha particles. The top panel shows the distribution of kinetic energies, $E_{kin}-rec$, as they are recorded by INDRA in the simulation, as a function of input $E_{kin}$ values. The bottom panel displays the corresponding FWHM width of the $E_{kin}-rec$ distributions as a function of $E_{kin}$.}
\label{sim1}
\end{figure}

 The widths of the observed
resonances are much larger than the intrinsic widths of the correspondent $^{12}$C states. For example, the FWHM (Full-Width-Half-Maximum) of the state at 9.65 MeV ($E_{k}\simeq$2.2 MeV) is observed to be about 1.0-1.1 MeV, to be compared to an intrinsic width of 34 keV for this state. This discrepancy  is mainly due to the finite angular resolution of the apparatus. We have performed a Monte Carlo simulation in order to estimate the effect of the INDRA detector on the resolution of the reconstructed excited states of  $^{12}$C decaying into three alpha particles. The excitation energy of $^{12}$C was sampled continuously between 0 and 15 MeV. The simulated events were filtered through the geometry and detector response of INDRA.  Fig.~\ref{sim1} shows the distribution of the three-body kinetic energy, $E_{kin}-rec$, as it is recorded by INDRA in the simulation, vs the input kinetic energy, $E_{kin}$. The average recorded kinetic energy, $E_{kin}-rec$, is slightly shifted towards higher values as compared to $E_{kin}$. Also, the width in the $E_{kin}-rec$ distribution increases with increasing $E_{kin}$. This is best shown on the bottom panel of Fig.~\ref{sim1} displaying the FWHM width of the $E_{kin}-rec$ distribution as a function of $E_{kin}$. At $E_{kin}$=2.2 MeV, which corresponds to the 9.65 MeV state, the FWHM is about 1.0 MeV, close enough to the measured one. From this study, it is clear that the position of the resonances is accurate but their widths are not.

Stimulated by the analysis technique described in Ref. \cite{cha95},
we modified the definition of the uncorrelated three particle yields
used in the denominator of Eq. \ref{eqfctcorr} by using a \textit{partial
event mixing} (PEM) technique. In particular, the open symbols on Fig. \ref{corr12c}
correspond to the correlation function obtained when the denominator
is constructed with two $\alpha$ particles taken from the same event
and the third one from a different event. The resonant peaks are still observed
in the correlation function, but with a reduced magnitude. This
reduction can be partly attributed to the presence of 2-body correlations
associated to sequential decays of $^{12}$C. Indeed, the decay of
these $^{12}$C states can proceed via the emission of $^{8}$Be+$\alpha$ pairs, 
with the very loosely bound $^{8}$Be nucleus subsequently decaying into two $\alpha$ particles. 
These second step $^{8}$Be decays are included in the numerator of Eq.~\ref{eqfctcorr}
but not in the denominator, if the latter is evaluated with the standard
event mixing technique. In contrast, when the PEM technique is used, two $\alpha$ particles in the 
denominator are still taken from the same event. In this case, some $\alpha-\alpha$ correlations due to 
secondary $^{8}$Be decays are kept in the denominator of Eq.~\ref{eqfctcorr}. Then, the sequential $^{8}$Be 
decay contributions contained in the numerator of Eq.~\ref{eqfctcorr} are partially cancelled out 
by calculating the ratio, $Y_{corr}/Y_{uncorr}$, and the magnitude of the correlation function peaks is reduced. 
Therefore, the observed attenuation of the peaks magnitude represents, by itself, an indication of the 
existence of $^{12}$C sequential decay, $^{12}$C$\rightarrow ^{8}$Be+$\alpha \rightarrow \alpha + \alpha + \alpha$. 

\begin{figure}
\centering
\includegraphics[scale=0.7]{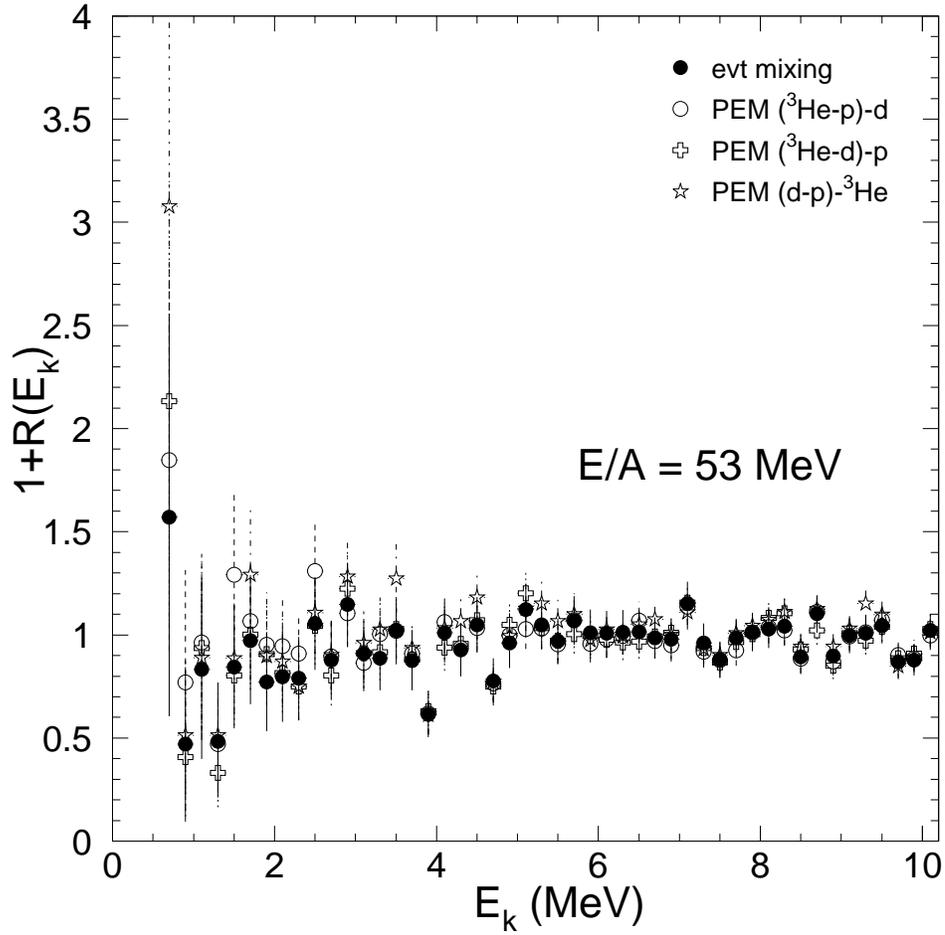}
\caption{$^{3}$He-$^{2}$H-$^{1}$H correlation functions from the $^{4}$He-$^{3}$He-$^{2}$H-$^{1}$H 
QP decay channel. Full dots: 1+R(E$_{k}$) constructed with the standard event mixing technique. 
Open symbols: 1+R(E$_{k}$) constructed with the PEM technique. Particles
within parentheses are taken from the same event.}
\label{corr6be}
\end{figure}

In order to test the validity of the method, we construct the correlation function for a case where no resonances are expected. For example, we select the four-body p-d-$^{3}$He-$\alpha$ exit channel events. The  $^{10}$C quasi-projectile could decay by first emitting $^4$He and $^6$Be. The $^6$Be nuclei is unbound but its states are not expected to decay into p-d-$^3$He\cite{niveau4,niveau7}. Therefore no peaks are expected in the correlation function.  Fig.~\ref{corr6be} shows the correlation function for the three particles constructed with the standard event mixing technique (full dots) and the PEM technique (open symbols). All combinations are free of correlations, the correlation functions are flat and average to unity as expected. 
 The enhancement of the correlation function observed at low E$_{k}$ values is not significant, mostly due to the very low statistics data collected in this low energy region. The observed slight tendency for 1+R(E$_{k}$) to be less than unity at E$_{k}<$2 MeV is also due to the suppression of the yields of detected events at low E$_{k}$ values, as it suggested by our Monte Carlo simulations. Difficulties in measuring R(E$_{k}$) at low E$_{k}$ are strongly due to the low angular resolution of INDRA.

\section{Multi-particle correlations and $^{10}$C excited states}

\begin{figure}
\centering
\includegraphics[scale=0.55]{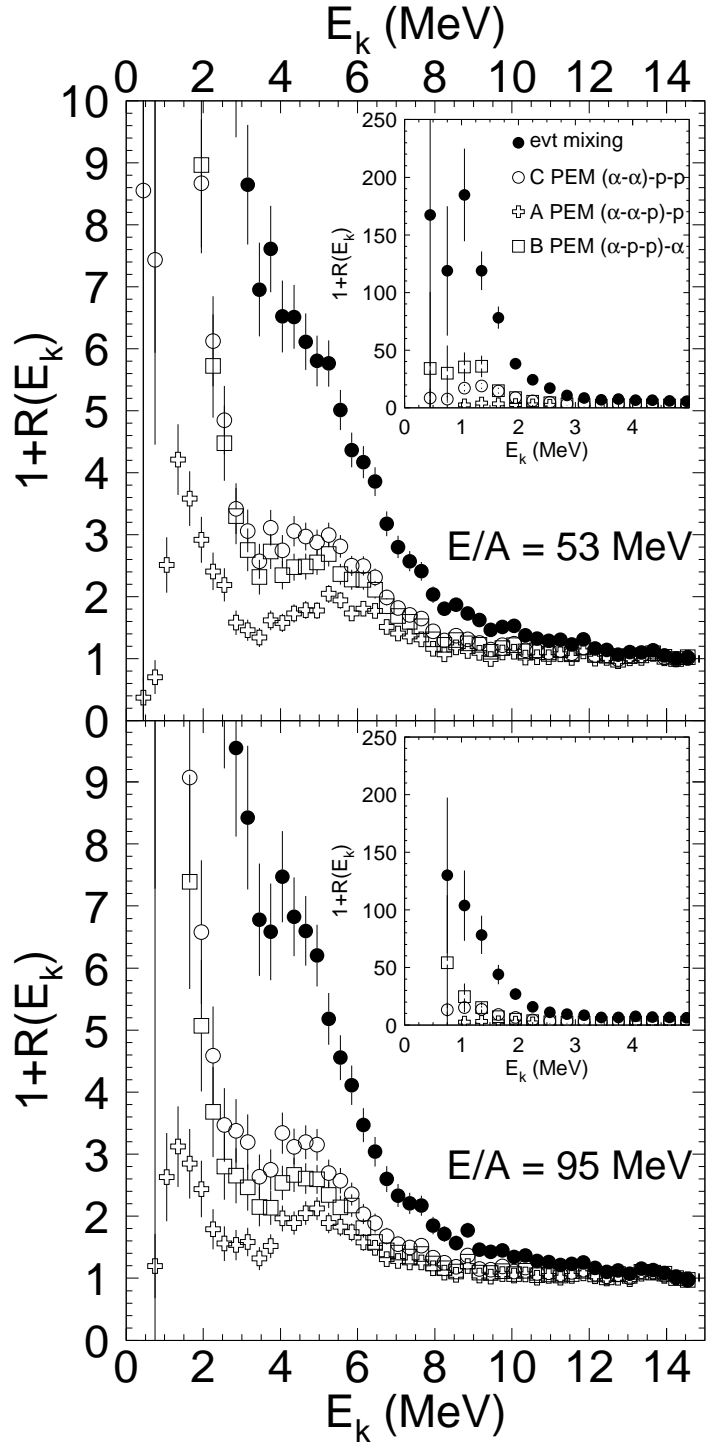}
\caption{2$\alpha$-2$p$ correlation functions at E/A=53 MeV (top panel) and 95 MeV (bottom panel). Solid symbols: standard 
event mixing technique. Open symbols: PEM techniques with two $\alpha$'s (circles), two $\alpha$'s and one proton (crosses) 
and one $\alpha$ and two protons (squares) taken from the same event (see text for details). Expanded views of the correlation 
functions for $E_{k}\leq$5 MeV are shown in the inset.}
\label{corr10c}
\end{figure}

We now turn to the study of $^{10}$C states. Fig.~\ref{corr10c} shows the four-particle 
2$\alpha$-2$p$ correlation function, $1+R(E_{k})$, constructed by using both the standard (closed symbols) 
and partial event mixing techniques (open symbols) already described in the case of 3$\alpha$ correlation functions. 
The top (bottom) panel refers to an incident energy of E/A=53
MeV (E/A=95 MeV). The 2$\alpha$-2$p$ correlation function constructed with the PEM technique (full dots) shows two broad peaks. The Q-value of the 2$p$+2$\alpha$ decay of $^{10}$C is -3.7 MeV (see the level scheme of $^{10}$C in ref~\cite{niveau8,niveau7}). The 
first peak at $E_{k}\approx$1.5 MeV can be associated to an overlap of $^{10}$C states at 
E$^{*}\approx$5.2-5.4 and 6.6 MeV. The second peak at $E_{k}\approx$5-5.5 MeV
corresponds to unknown states around E$^{*}\approx$9 MeV \cite{Wan93}. The location of
these peaks is independent of the beam energy. The PEM explores the following decays sequences: A) $^{10}$C$\rightarrow^{9}$B+p with the $^{9}$B nucleus further undergoing a decay 
$^{9}$B$\rightarrow\alpha$+$\alpha+$p (open crosses); B) $^{10}$C$\rightarrow^{6}$Be+$\alpha$ with 
the $^{6}$Be nucleus further decaying into $\alpha$+p+p  (open squares); 
C) $^{10}$C$\rightarrow^{8}$Be+p+p with the $^{8}$Be nucleus further decaying 
into $\alpha$+$\alpha$  (open circles). 

All PEM correlation functions display an attenuation of the peak magnitude. In the case of the states of $^{10}$C observed at E$_{k}\approx$5-5.5 MeV, the attenuation obtained with the PEM techniques 
is less pronounced as compared to the case of the lower lying state at E$_{k}\approx$1.5 MeV. 
This observation suggests that direct four body decays (2$p$+2$\alpha$) without passing 
through any intermediate unbound state are more likely for the states around E$^{*}\approx$9 MeV 
than for the state at E$^{*}\approx$5.2-5.4 MeV. 
Among the studied sequential decay modes, sequence A induces a peak 
magnitude attenuation of the order of 30$\%$ with respect to its original value (see open crosses).  
Decay sequences B and C seem to provide comparable contributions to the decay of the studied $^{10}$C state, 
within statistical uncertainties, reducing the peak magnitude to about 35-40$\%$ of its original value. A slight 
systematic preference for the B sequence as compared to the C sequence is observed. In general, for both the first and second group of peaks we observe a preference of $^{10}$C to decay through the production of $^{9}$B, as compared to a decay proceeding through the  $^{6}$Be system. 

\begin{figure}
\centering
\includegraphics[scale=0.7]{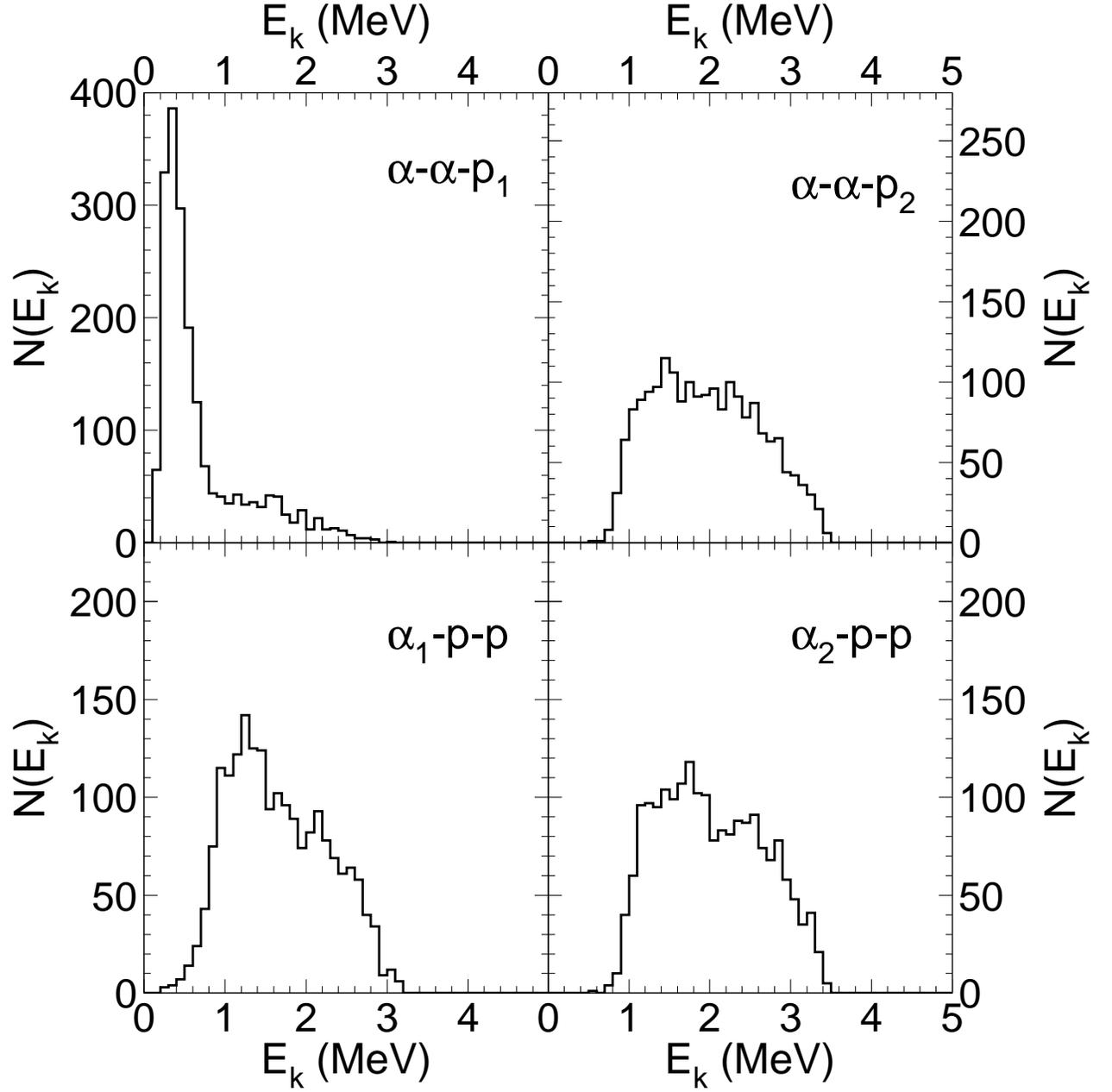}
\caption{Top panels: $\alpha-\alpha-p$ kinetic energy spectra constructed with the slowest ($p_{1}$, left panel) and the 
fastest ($p_{2}$, right panel) proton. Bottom panel: $\alpha-p-p$ kinetic energy spectra constructed with the slowest 
($\alpha_{1}$, left panel) and the fastest ($\alpha_{2}$, right panel) $\alpha$ particle. The data correspond to an 
incident energy of E/A=53 MeV.} 
\label{seq10c}
\end{figure}

In order to confirm these conclusions about  $^{10}$C sequential decay modes, we study three-particle coincidence spectra for events with 1 MeV$\leq$ E$_{k}\leq$ 3.5 MeV, corresponding to the first peak on Fig.~\ref{corr10c}. On Fig.~\ref{seq10c} (top panels) we show 
$\alpha-\alpha-p_{1}$ and $\alpha-\alpha-p_{2}$ total kinetic energy spectra, $N(E_{k})$, where the total kinetic 
energy, $E_{k}$, is calculated in the three-body reference frame. In particular, we choose the slowest ($p_{1}$) or 
the fastest ($p_{2}$) proton in the $^{10}$C reference frame. Similar spectra are constructed using the $\alpha_{1}-p-p$ 
and $\alpha_{2}-p-p$ coincidences (bottom panels on Fig.~\ref{seq10c}), where $\alpha_{1}$ and $\alpha_{2}$ particles are chosen 
using the same criteria as  $p_{1}$ and $p_{2}$. These spectra of Fig.~\ref{seq10c} are expected to contain information about 
possible intermediate states in $^{9}$B and $^{6}$Be nuclei produced during  $^{10}$C decay sequences A and B 
described above. If $^{10}$C$^{*}$ did not decay through a definite sequence,
the $N\left(E_{k}\right)$ distributions would show no signatures of nuclear resonant decays. In contrast, we observe 
a peak at $E_{k}\approx$0.2 MeV in the $\alpha-\alpha-p_{1}$ spectrum, corresponding to the decay of $^{9}$B$_{g.s.}$ 
(decay sequence A), and a peak at $E_{k}\approx$1.4 MeV in the $\alpha_{1}-p-p$ spectrum, corresponding to the 
decay of $^{6}$Be$_{g.s.}$ (decay sequence B). In the spectra $\alpha-\alpha-p_{2}$ and $\alpha_{2}-p-p$ (right panels 
on Fig.~\ref{seq10c}) we observe broad peaks possibly associated to the decay of higher $^{9}$B and $^{6}$Be excited states. 
These observations seem to confirm our conclusions deduced from the PEM analysis shown on Fig.~\ref{corr10c}. The results obtained with our analysis techniques provide an experimental evidence of sequential decay modes of unbound states of $^{10}$C exotic nuclei.  Furthermore,  our estimates of contributions from these different sequential modes 
have been recently confirmed by a dedicated experiment performed with a high resolution detector array and by using a different analysis techniques \cite{char07}.
 
In conclusion, the existence of sequential decay modes of $^{10}$C states proceeding through the production of intermediate loosely bound $^{6}$Be, $^{8}$Be and $^{9}$B nuclei has been explored. 
The analysis method, based on pure and modified event mixing techniques, has the advantage of not containing possible 
model-dependencies induced by the use of user-constructed simulations to build the uncorrelated background. Furthermore, 
the modified event mixing techniques allow us to semi-quantitatively estimate the 
relative contributions from different decay channels. A preference for a decay 
path producing unstable $^{9}$B nuclei is found for the states at 5.2-5.4 MeV in $^{10}$C. 
The obtained results are also in good agreement with those extracted with high resolution 
experiments \cite{char07}.
Our results confirm that heavy-ion collisions 
at intermediate energies can be used as a 
spectroscopic tool to access simultaneously important spectroscopic properties of several unstable nuclei all  produced in one single collision. This perspective encourages an extension of similar  studies to 
collisions induced by radioactive beams where more exotic nuclear species are expected to 
be produced.

\section{Acknowledgments}
We thank the staff of the GANIL Accelerator facility
for their support during the experiment. This work was supported by
four France institutions : Le Commissariat \`{a} l'Energie Atomique,
Le Centre National de la Recherche Scientifique, Le Minist\`{e}re
de l'Education Nationale, and le Conseil R\'egional de Basse Normandie.
This work was also supported by the Natural Sciences and Engineering
Research Council of Canada, the Fonds pour la Formation de Chercheurs
et l'Aide \`{a} la Recherche du Qu\'{e}bec.

% \appendix

% \section{}
% \label{}

%\begin{thebibliography}{00}

%\bibliography{reference}


% notes:

% \bibitem{label2}
% If there is a note, it should come last:
% \bibitem{label3} \note
% \end{subbibitems}

%\end{thebibliography}

\newpage

\newpage

\newpage

\newpage

\newpage

\newpage

\newpage

\newpage

\end{document}